\documentstyle[aps,prb,epsf]{revtex}
\begin{document}
\twocolumn[\hsize\textwidth\columnwidth\hsize\csname@twocolumnfalse\endcsname
\title{Plasma resonance at low magnetic fields as a probe of vortex 
line meandering in layered superconductors }

\author{L.\ N.\ Bulaevskii$^a$, A.\ E.\ Koshelev$^b$, V.\ M.\ Vinokur$^b$, and 
M.\ P.\  Maley$^a$}

\address{$^a$ Los Alamos National Laboratory, Los Alamos, NM 87545 }
\address{$^b$ Materials Science Division, Argonne National Laboratory, 
Argonne, IL 60439}

\date{\today}

\maketitle
\begin{abstract}
We consider the magnetic field dependence of the plasma resonance
frequency in pristine and in irradiated Bi$_2$Sr$_2$CaCu$_2$O$_8$
crystals near $T_c$.  At low magnetic fields 
we relate linear in field
corrections to the plasma frequency to the average distance between
the pancake vortices in the neighboring layers (wandering length).
We calculate the wandering length in the case of thermal wiggling of vortex
lines, taking into account  both Josephson and magnetic interlayer coupling 
of pancakes.  
Analyzing experimental data, we found that (i) the wandering length
becomes comparable with the London penetration depth near T$_{c}$ and (ii)
at small melting fields ($< 20$ G) the wandering length does not change
much at the melting transition.  This shows existence of the line liquid phase in this 
field range. We also found that pinning by columnar defects affects
weakly the field dependence of the plasma resonance frequency near
$T_c$.
\end{abstract}

\vskip.2pc] 
\narrowtext 

Josephson plasma resonance (JPR) measurements in highly anisotropic layered 
superconductors provide unique information on the interlayer Josephson 
coupling and on the effect of pancake vortices on this coupling.  
The squared c-axis plasma resonance frequency, $\omega_p^2$, is
proportional to the average interlayer Josephson energy,
$\omega_p^2\propto J_0\langle\cos\varphi_{n,n+1}({\bf r}) \rangle$,
where $J_0$ is the Josephson critical current, $\varphi_{n,n+1}({\bf
r})$ is the gauge-invariant phase difference between layers $n$ and
$n+1$ and ${\bf r}$ is the in-plane coordinate.  Here $\langle \ldots
\rangle$ means average over thermal disorder and pinning.  Thermal
fluctuations and uncorrelated pinning lead to misalignment of pancake
vortices induced by the magnetic field applied along the $c$ axis. 
Misalignment results in nonzero phase difference and in the
suppression of Josephson coupling and plasma
frequency.\cite{bul} Thus, the $\omega_{p}$ dependence on the $c$-axis
magnetic field measures the $c$-axis correlations of pancakes in the
vortex state.

The JPR measurements performed in the liquid vortex phase at
relatively high magnetic fields, $B>B_J=\Phi_0/\lambda_J^2$ revealed
that the plasma frequency drops approximately as
$1/\sqrt{B}$.\cite{matsuda} Here $\lambda_J= \gamma s$ is the
Josephson length, $\gamma$ is the anisotropy ratio and $s$ is the
interlayer distance.  The above dependence is characteristic for the
pancake liquid weakly correlated along the $c$
axis.\cite{kosh,kbmPRL98} Here a pancake in a given layer is shifted
by a distance of the order of vortex spacing $a=(\Phi_0/B)^{1/2}\ll
\lambda_{J}$ from the nearest pancake in the neighboring layer.  Thus
at high fields many pancake vortices contribute to the suppression of
the phase difference at a given point ${\bf r}$, because $\lambda_J$
determines the decay length for the phase difference induced by
misaligned pancakes of a given vortex line.\cite{blk} In contrast, in
the vortex solid a lattice of vortex lines forms as shown by neutron
scattering- and $\mu ^{+}$SR data.\cite{Cubitt} JPR measurements in
Bi$_2$Sr$_2$CaCu$_2$O$_{8-\delta}$ (Bi-2212) crystals
\cite{sh,shib,mats} have shown that in the fields above 20 Oe the
interlayer phase coherence changes drastically at the transition line
[$\langle\cos\varphi_{n,n+1}({\bf r}) \rangle$ jumps], implying the
decoupling nature of the first-order melting transition, see
discussion in Ref.\onlinecite{BlatterPRB96}.

In this Letter we focus on the low magnetic fields, $B< B_J$, near $T_c$ 
regime.
Here the intervortex distance is much larger than $\lambda_J$, and the
Josephson coupling in the region occupied by a given vortex is not
suppressed by other vortices.  In such a single vortex regime the
Josephson energy increases linearly with displacements of nearest
pancakes in neighboring layers, when $|{\bf u}_{n,n+1}|\equiv|{\bf
r}_n-{\bf r}_{n+1}|>\lambda_J$, see Ref.~\onlinecite{blk} (here ${\bf
r}_n$ is the coordinate of a pancake in the layer $n$).  This leads to
the confinement of pancakes in neighboring layers, and $c$-axis
correlated pancakes (i.e. vortex lines) may be preserved above the
melting transition.  We show that it is the linear decrease of
$\omega_p^2$ with $B$ that characterizes such a vortex state.  The
linear dependence was observed experimentally in
Refs.~\onlinecite{sh,shib,mats} in both solid and liquid vortex states
in Bi-2212 crystals in fields below 20 Oe near $T_c$ providing
evidence for a line structure of the vortex liquid state at low
fields.

We calculate the plasma frequency at low magnetic fields $B$, and near 
$T_c$ assuming that only vortices induced by the applied magnetic field 
${\bf B}\parallel c$ contribute to the field dependence of $\omega_p$.  
 We, thus, ignore the contribution of thermally excited vortices and 
antivortices to the field dependence of the plasma frequency.

The JPR absorption is described by a simplified equation \cite{bdmb} for 
small oscillations of the phase difference $\varphi_{n,n+1}^{\prime }({\bf 
r},\omega)$ induced by an external microwave electric field with the 
amplitude ${\cal D}$ and the frequency $\omega$ applied along the $c$ axis:
\begin{equation}
\left[ \frac{\omega(\omega+i\Gamma_c)}{\omega_{0}^{2}}-1+\lambda 
_{J}^{2}\hat{L}\nabla_{{\bf r}} ^{2}-{\cal V}_n({\bf r})\right] 
\varphi _{n,n+1}^{\prime }=-\frac{ i\omega {\cal D}}{4\pi J_{0}}.
\label{DynEq}
\end{equation}
Here ${\cal V}_n({\bf r})=\cos\varphi_{n,n+1}({\bf r})-1$ is the
effective potential, $\varphi_{n,n+1}({\bf r})$ is the phase
difference induced by vortices misaligned due to thermal fluctuations
and pinning in the absence of a microwave field.  In Eq.~(\ref{DynEq})
we neglect the time variations of $\varphi_{n,n+1}({\bf r},t)$ because
the plasma frequency is much higher than the characteristic
frequencies of vortex fluctuations, see below.  Further, $\omega
_{0}(T)=c/\sqrt{\epsilon _{0}}\lambda _{c}(T)$ is the zero field
plasma frequency \cite{phasefluct}, $\epsilon_{0}$ is the high
frequency dielectric constant, $\lambda_{ab}$ and $\lambda_{c}=\gamma
\lambda_{ab}$ are the components of the London penetration depth,
$E_{J}=E_{0}/\lambda_J^2$ is the Josephson energy per unit area and
$E_{0}=s\Phi _{0}^{2}/16\pi^{3}\lambda _{ab}^{2}$ is the
characteristic pancake energy.  The inductive matrix $\hat{L}$ is
defined as $\hat{L}A_{n}=\sum_{m}L_{nm}A_{m}$ with $ L_{nm}\approx
(\lambda _{ab}/2s)\exp\left( -|n-m|s/\lambda _{ab}\right)$.  The
parameter $\Gamma_c=4\pi\sigma_c/\epsilon_0$ describes dissipation due
to quasiparticles.  Here $\sigma_c$ is the $c$-axis quasiparticle
conductivity in the superconducting state.  Practically it coincides
with conductivity right above $T_c$.

The absorption in the uniform AC electric field is defined by the
imaginary part of the inverse dielectric function
\begin{equation}
{\rm Im}\frac{1}{\epsilon(\omega)}=\frac{1}{\epsilon_0}\sum_{\alpha n}
\int d{\bf r}\frac{\langle \Psi_{\alpha n}^*(0)\Psi_{\alpha n}({\bf r})
\rangle\omega^3
\Gamma_c}{(\omega^2-\omega_{\alpha}^2)^2+\omega^2\Gamma_c^2},
\label{ab}
\end{equation}
where $E_{\alpha}=1-\omega_{\alpha}^2/\omega_0^2$ and $\Psi_{\alpha n}({\bf 
r})$ are the eigenvalues and eigenfunctions of the operator $-\lambda 
_{J}^{2}\hat{L}\nabla_{{\bf r}} ^{2}+{\cal V}_n({\bf r})$.

\begin{figure}
\epsfxsize=3.1in \epsffile{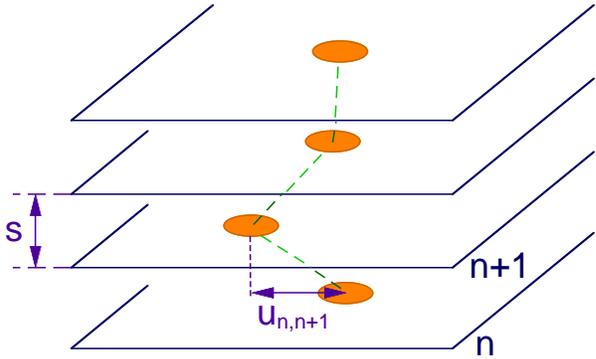} 
\caption{Meandering of pancakes 
along the vortex line in the single vortex regime
at low magnetic fields in highly anisotropic layered superconductors.}
\label{FigMeand}
\end{figure}

Consider magnetic fields $B\ll \Phi_0/4\pi\lambda_{ab}^2, B_J$ (single 
vortex regime).  The phase difference near a given vortex line 
is induced by displacements 
of pancakes in neighboring layers along this vortex line (see  
Fig.\ \ref{FigMeand}).  The potential 
${\cal V}_{n}({\bf r})$ at distances $u_{n,n+1}\ll r\ll \lambda _{J}$ is 
determined by the phase difference produced by nearest pancakes in 
neighboring layers $n$ and $n+1$ relatively displaced at distance ${\bf 
u}_{n,n+1}$:
\begin{equation}
\varphi _{n,n+1}({\bf r})\approx[{\bf r}\times {\bf u}_{n,n+1}]/r^{2}.
\label{phi}
\end{equation}
At large distances $r\gg\lambda_J$ the potential drops exponentially
and at small distances $r\lesssim u_{n,n+1}$ it tends to a constant. 
This potential is attractive and there are localized and delocalized
states.  At $a\gg\lambda_J,u_{n,n+1}$ main contribution to absorption
is coming from
the most homogeneous delocalized state.  Such a state determines the
center of JPR line $\omega_p(B,T)$.  Other states lead to
inhomogeneous line broadening in addition to broadening caused by
quasiparticle dissipation described by $\Gamma_c$.  The latter
mechanism of broadening dominates at low magnetic fields near $T_c$.

The strength of the potential with respect to the kinetic term, $\lambda 
_{J}^{2}\hat{L}\nabla_{{\bf r}} ^{2}$, is characterized by the 
dimensionless parameter $r_w^2/\lambda_J^2$ which we assume to be small in 
the following calculations.  Here $r_{w}$ is the elemental wandering length 
of vortex lines, $r_{w}^{2}\equiv \langle {\bf u}_{n,n+1}^{2}\rangle $.  
Then we use perturbation theory with respect to the potential ${\cal 
V}_{n}({\bf r})$ to find the energy of the most uniform delocalized state.  
The unperturbed wave function of this state is given by a constant.  The 
first order correction to the energy of this most homogeneous delocalized 
state is given by the space average of the potential, averaging is over 
${\bf r}$ and $n$.  This space average is equivalent to the thermal average 
of $\cos\varphi_{n,n+1}({\bf r})-1\approx -\varphi_{n,n+1}^2({\bf r})/2$.  
Using Eq.~(\ref{phi}) we obtain a simple relation connecting field-induced 
suppression of the plasma frequency $\omega_p(B,T)$ with $r_{w}$ for the 
case $r_{w}<\lambda _{J}$:
\begin{equation}
\frac{\omega_0^2(T)-\omega_p^2(B,T)}{\omega_0^2(T)}\approx\frac{
\langle \varphi_{n,n+1}^2({\bf r})\rangle}{2}\approx \frac{\pi Br_{w}^{2}}{
2\Phi _{0}}\ln \frac{\lambda _{J}}{r_{w}}.  \label{Cosrw}
\end{equation}
The relation (\ref{Cosrw}) is very general and does not depend on the 
mechanism of the vortex wandering.  It allows one to extract $ r_{w}^{2}$ 
from the plasma resonance measurements.  The field dependence of the 
resonance temperature, $T_{r}(B)$, is determined by the equation $\omega 
_{p}^{2}(B,T_{r})=\omega ^{2}$.  According to Eq.~(\ref {Cosrw}) this gives 
a linear dependence at small fields, $T_{r}(B)\approx 
T_{r}(0)+(dT_{r}/dB)B$, and $r_{w}^{2}$ is directly related to the slope of 
this dependence
\begin{equation}
r_{w}^{2}\ln \frac{\lambda _{J}}{r_{w}}=\frac{2\Phi _{0}}{\pi }
\frac{d\ln \omega _{0}^{2}(T)}{dT}
\left(\frac{dT_{r}}{dB}\right)_{B\rightarrow 0}.  
\label{rwTexp}
\end{equation}
We now calculate $r_{w}^{2}$ when wandering of the vortex lines is caused by
thermal fluctuations. In the single vortex regime $r_{w}^{2}$ is determined
by the wandering energy consisting of the Josephson and magnetic
contributions, 
\begin{equation}
{\cal F}_{w}\approx \frac{s}{2}\sum_{n}\left[ \varepsilon _{1J}
\left(\frac{ {\bf r}_{n+1}-{\bf r}_{n}}{s}\right) ^{2}+w_{M}{\bf r}
_{n}^{2}\right], 
\end{equation} 
where $\varepsilon _{1J}\approx (\Phi _{0}^{2}/(4\pi \lambda_{c})^{2}) \ln 
(\lambda _{J}/r_{w})$ is the line tension due to the Josephson coupling and 
$w_{M}\approx (\Phi _{0}^{2}/ (4\pi \lambda _{ab}^{2})^{2})\ln (\lambda 
_{ab}/r_{w})$ is the effective cage potential, which appears due to 
strongly nonlocal magnetic interactions between pancake vortices in 
different layers.\cite{kkPRB93,BlatterPRB96}  Assuming Gaussian 
fluctuations we have
\begin{eqnarray}
&&r_{wT}^{2}=\int \frac{dq_{z}}{2\pi }\frac{4T(1-\cos q_{z}s)}{2\left(
\varepsilon _{1J}/s^{2}\right) \left( 1-\cos q_{z}s\right) +w_{M}}
=\frac{2sT}{\varepsilon _{1J}}f(\zeta), \nonumber \\
&&f(\zeta )=\frac{\zeta }{1+\zeta +\sqrt{1+\zeta }}, 
\label{rwT} 
\end{eqnarray}
where the parameter $\zeta (T)=4\lambda _{ab}^{2}(T)/\lambda _{J}^{2}$ describes 
the relative roles of the Josephson and magnetic interactions. Substituting
this result into Eq.~(\ref{Cosrw}) we obtain 
\begin{equation}
\frac{\omega_0^2(T)-\omega_p^2(B,T)}{\omega_0^2(T)}
\approx \frac{B}{B_{0} }f(\zeta )
\label{ompB}
\end{equation}
with $B_{0}=\Phi _{0}^{3}/16\pi ^{3}\lambda _{c}^{2}sT=B_{J}(E_{0}/T)$.
We stress that this result of the single vortex regime is valid in both 
solid and liquid vortex states for $B\ll B_{J}$, because in this field 
range wandering of lines at short scales does not change much at the 
melting point.  The difference between these states appears only in the 
second order in the magnetic field.  At $B=0$ the resonance occurs at the 
temperature $T_r(\omega)$. 
The slope of the curve $B_{r}(T)$ at small $B_{r}$ is
\begin{equation}
\frac{dB_{r}}{dT}= 
\frac{1}{f(\zeta_{r})}\left(\frac{dB_{0}}{dT}\right)_{T=T_{r}},\ \ \ 
\zeta _{r}=\frac{4\lambda _{ab}^{2}(T_{r})}{\lambda _{J}^{2}}.  
\label{sl}
\end{equation}
The crossover region from the line liquid, $r_w^2\ll a^2$, to the pancake 
liquid, where $r_w^2\approx a^2$, takes place at the magnetic field 
$B\approx {\rm min}[\pi B_0/2f(\zeta), B_J]$.

\begin{figure}
\epsfxsize=3.2in \epsffile{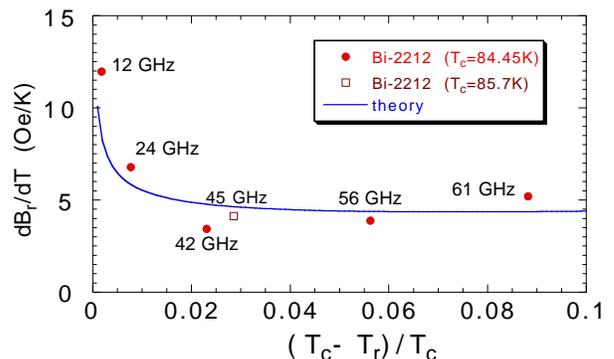} 
\caption{Comparison of the 
experimental temperature dependence of $dB_{r}/dT$ obtained in Ref.\ 
\protect\onlinecite{sh} using different microwave frequencies with 
theoretical dependence from Eq.\ \protect (\ref{sl}) (see text).  We also 
show data point from Ref.\ \protect \onlinecite{mats} obtained for Bi-2212 
with slightly different $T_{c}$.  }
\label{FigdBrdT}
\end{figure}

%
To compare our calculation with experiment we plot in Fig.~\ref{FigdBrdT} 
dependence of $dB_{r}/dT$ vs reduced resonance temperature at $B=0$, 
$t_{r}=(T_{c}-T_{r})/T_{c}$, obtained in Ref.~\onlinecite{sh,shib} using 
different microwave frequencies (shown in the plot) for Bi-2212 with 
$T_{c}=84.45$ K. We also show data point obtained by Matsuda {\em et al.} 
\cite{mats} for Bi-2212 with close $T_{c}$, $T_{c}=85.7$ K. To calculate 
the dependence $dB_{r}/dT$ from Eq.\ (\ref{sl}) we need dependencies 
$\lambda_{c}(T)$ and $\lambda_{ab}(T)$, which determine dependencies 
$B_{0}(T)$ and $\zeta(T)$.  $\lambda_{c}(T)$ was found from temperature 
dependence of $\omega_{0}$ at zero field, which we fit as 
$\omega_{0}(t)/2\pi \approx (133.5\ {\rm GHz})t^{0.32}$, and taking 
$\epsilon_{c}=11$.  Matsuda {\em et al.} \cite{mats} obtained similar 
temperature dependence of $\omega_{0}$ using direct frequency scan.  
$\lambda_{ab}(T)$ was obtained assuming temperature independent $\gamma$, 
which we adjusted to obtain the best agreement between the theoretical and 
experimental curves giving $\gamma=480$.  Nonmonotonic temperature 
dependence of $dB/dT_r$ arises from competition between two factors in Eq.\ 
(\ref{sl}): increase at low temperatures is due to the factor 
$1/f(\zeta)\propto \lambda_{ab}^{-2}$ at $\zeta \ll 1$, and increase at 
temperatures close to $T_{c}$ is due to nonlinearity of the dependence 
$\lambda_{c}^{-2}(T)$, which leads to the divergency of $dB_{0}/dT$ at 
$T\rightarrow T_{c}$.  As one can see from the plot, our theory describes 
satisfactorily the field dependence of $\omega_{p}$ not very close to 
$T_{c}$, i.e., in the region where the critical fluctuations are not very 
strong.  The region of critical fluctuations is beyond applicability of our 
theory.

We can now check validity of our approximations: the static approximation 
for the potential in Eq.~(\ref{DynEq}) and the perturbation theory with 
respect to the potential.  The maximum frequency of vortex fluctuations in 
the single vortex regime is $\omega_{fl}\approx \varepsilon _{1J}/s^2\eta$, 
where the vortex viscosity $\eta$ estimated from flux flow resistivity is 
in the interval $10^{-6}-10^{-7}$ g/cm$\cdot$s, see Ref.~\onlinecite{bm}.  
This gives $\omega_{fl}/\omega_p\approx 
\Phi_0^2\sqrt{\epsilon_0}/16\pi^2s^2\lambda_cc\eta \lesssim 0.1$ near 
$T_c$.  Condition $r_w^2/\lambda_J^2\lesssim 1$ for applicability of the 
perturbation theory is satisfied at $t> 0.02$.

Near $T_c$ The experimental curve $T_r(B)$ shows no change of slope when it crosses 
the melting line, $H_m\approx 1.3$[Oe/K]$(T_c-T)$.\cite{sh}  This gives 
evidence that the parameter $r_w^2$ does not change much at melting.  Using 
Eq.~(\ref{rwTexp}) we estimate that, near melting line $r_w^2/a^2\lesssim 
0.1$ at temperatures studied.  This confirms the line structure of the 
vortex liquid above the melting line at low fields, though vortex lines 
wander over extended distances already in the solid phase due to high 
temperatures, see Fig.~\ref{FigMeand}.  The estimated 
value, $r_w\approx 1$ $\mu$m at 77 K, is comparable with both $\lambda_J$ 
and $\lambda_{ab}$.  In Bi-2212 crystals near optimal doping crossover to a 
pancake liquid occurs in the field interval $\approx 10- 15$ Oe.  
In less anisotropic high-T$_{c}$ materials
one anticipates a larger region of the line liquid on the vortex phase
diagram.

Next we calculate the effect of columnar defects (CDs) on the parameter $ 
r_{w}^{2}$ at high temperatures.  Columnar defects always straighten vortex 
lines and reduce $r_{w}$.  At low temperatures this effect is very strong.  
Each vortex line is localized near one defect and its wandering is much 
smaller than in an unirradiated superconductor.  At high temperatures lines 
start to distribute over a large number of defects and the effect of CDs 
progressively decreases.  We consider the extreme case of very high 
temperatures when the effect of defects can be considered within 
perturbation theory.  This approach is applicable at temperatures higher 
than the pinning energy of pancakes by CD, $T>\pi E_0 \ln(b/\xi_{ab})$, 
where $b$ is CD radius and $\xi_{ab}$ is the superconducting correlation 
length.  This situation corresponds to experiments.\cite{sh,shib,mats}

The free energy functional is 
\begin{equation}
{\cal F}={\cal F}_{w}+{\cal F}_{p},\ \ \ {\cal F}_{p}=\sum_{n}U({\bf r}_{n}),
\label{fun}
\end{equation}
where $U({\bf r})=\sum_{i}V({\bf r}-{\bf R}
_{i})$ is the pinning potential of CDs, ${\bf R}_{i}$ are the positions of 
CDs, and $V({\bf r})=\pi E_{0}\ln
(1-b^{2}/r^{2})\exp (-r/\lambda _{ab})$ is the pinning potential of the
individual CD.\cite{shm} Expanding with respect to disorder up to 
second order terms we obtain the correction, $(-r_{wD}^{2})$, to the zero order
term, $r_{wT}^{2}$, due to pinning by CDs: 
\begin{eqnarray}
&&-r_{wD}^{2} =r_{w}^{2}-r_{wT}^{2}  \label{fi} \\
&&=-\frac{s}{T}\frac{\partial }{\partial \varepsilon _{1J}}\sum_{m}\left[
\left\langle K({\bf r}_{m}-{\bf r}_{0})\right\rangle _{0}-\left\langle
\left\langle K({\bf r}_{0}^{\prime }-{\bf r}_{0})\right\rangle
_{0}\right\rangle _{0}^{\prime }\right],  \nonumber
\end{eqnarray}
where $\left\langle \ldots \right\rangle _{0}$ stands for statistical
average for a system without disorder and $K({\bf r})$ is the correlation
function of disorder, \begin{equation}
K({\bf r}^{\prime }-{\bf r})=\langle U({\bf r}^{\prime })U({\bf r})\rangle
_{D}-\langle U({\bf r})\rangle _{D}^{2}.
\end{equation}
When $r_{w}$ is much larger than the distance between columns we can
transform Eq.~(\ref{fi}) to a simpler form 
\begin{equation}
r_{wD}^{2}=\frac{2sK_{0}}{T}\frac{\partial }{\partial \varepsilon _{1J}}
\sum_{m}\left[ \frac{2}{\left\langle ({\bf r}_{m}-{\bf r}_{0})^{2}\right
\rangle _{0}}-\frac{1}{\left\langle {\bf r}_{0}^{2}\right\rangle _{0}}
\right],   \label{rwD1}
\end{equation}
where $K_{0}\equiv \int d{\bf r}K({\bf r})=n_{\phi }\left[ \pi
^{2}b^{2}E_{0}\ln(\lambda_{ab}/b)\right]^{2}$. Here $n_{\phi }$ is the
concentration of CDs. The correction is determined
by the lateral line displacement $\left\langle ({\bf r}_{m}-{\bf r}
_{0})^{2}\right\rangle _{0}$ which we calculate as 
\begin{eqnarray}
&&\left\langle ({\bf r}_{m}-{\bf r}_{0})^{2}\right\rangle _{0} =\int \frac{
dq_{z}}{2\pi }\frac{4T(1-\cos mq_{z}s)}{2\left( \varepsilon _{1J}/s^{2}\right)
\left( 1-\cos q_{z}s\right) +w_{M}} \nonumber \\
&&=\frac{4sT}{\varepsilon _{1J}}\frac{v\left( 1-v^{m}\right) }{\left(
1-v^{2}\right) },\;\;\;v=\frac{\zeta +2-2\sqrt{\zeta +1}}{\zeta}.    
\label{rm2}
\end{eqnarray}
Combining Eqs.~(\ref{rwD1}) and (\ref{rm2}) we finally obtain for $r_{wD}^{2}$
\begin{equation}
r_{wD}^{2}=\frac{2K_{0}}{T^{2}}f_{D}(v).
\end{equation}
Here the dimensionless function \[
f_{D}(v)\equiv \frac{\left( 1-v\right) ^{3}}{1+v}\frac{d}{dv}\left( \frac{1+v
}{1-v}\sum_{n=1}^{\infty }\frac{1}{v^{-n}-1}\right) \]
has the limits $f_{D}(0)=1$ and $f_{D}(v)\approx 2.16-2\ln (1-v)$ 
at $v\rightarrow 1$. Comparing this equation with Eq.~(\ref{rwT}) we
obtain \begin{equation}
\frac{r_{wD}^{2}}{r_{wT}^{2}}\approx \frac{\pi ^{2}n_{\phi }b^{4}}{\lambda
_{J}^{2}}\left( \ln ^{2}\frac{\lambda _{ab}}{b}\right) \left( \frac{E_{0}}{T}
\right) ^{3}\ln \frac{\lambda _{J}}{b}.
\end{equation}
In the crystals studied $\lambda _{J}\approx 1$ $\mu $m, $\lambda
_{ab}(0)\approx 2000$ \AA , $b\approx 70$ \AA . For the temperature range
$T>77$ K explored in Refs.\ \onlinecite{shib,mats} by the plasma resonance 
this gives a very small correction ($r_{wD}^{2}/r_{wT}^{2}\approx 10^{-3}$).
Therefore, the effect of pinning by CDs on the field dependence 
of the plasma frequency near $T_c$ is negligible.  

It was found in Refs.~\onlinecite{shib,mats} that after irradiation
with the matching field $B_{\phi }=\Phi_0n_{\phi}\approx 1$ T the
value $(dB/dT_r)_{B\rightarrow 0}$ near $T_{c}$ increases about two
times in comparison with that in pristine crystals.  As was estimated
above, pinning due to CDs cannot give such a strong effect.  One may
think that irradiation reduces the value of the anisotropy parameter
$\gamma $, probably due to damage of the crystal structure around the
heavy ion tracks.  This assumption is consistent with recent
measurements of the Josephson current in irradiated Bi-2212 mesas by
Yurgens {\it et al.} \cite{yurgens} It was found that irradiation
approximately doubles the Josephson current at zero field.

In conclusion, we have calculated the field dependence of the JPR frequency 
in the single vortex regime at low magnetic fields near $ T_{c}$ and 
demonstrated that the JPR provides a direct probe for meandering of 
individual lines.  We have shown that the JPR data in highly anisotropic 
Bi-2212 crystals give evidence that at high magnetic fields $B\gg B_{J}$ 
pancakes are uncorrelated along the $c$ axis in the vortex liquid (pancake 
liquid), while at lower fields, $B\leq B_{J}$, pancakes form vortex lines 
(line liquid).  These lines, however, strongly meander in both solid and 
liquid vortex states due to thermal fluctuations.  We have shown also that 
JPR data provide evidence that irradiation by heavy ions causes a 
significant decrease of the effective anisotropy.

The authors thank Y. Matsuda, M. Gaifullin, T. Tamegai, and T. Shibauchi 
for providing their experimental data prior publication.  This work was 
supported by the NSF Office of the Science and Technology Center under 
contract No.\ DMR-91-20000 and by the U.\ S.\ DOE, BES-Materials Sciences, 
under contract No.\ W-31-109-ENG-38.  Work in Los Alamos is supported by 
the U.\ S.\ DOE.


\begin{references}

\bibitem{bul}L.\ N.\ Bulaevskii, M.\ P.\ Maley, and M.\ Tachiki, Phys.\ 
Rev.\ Lett. {\bf 74}, (801) (1995).
\bibitem{matsuda}Y.\ Matsuda, {\it et al.}, 
Phys.\ Rev.\ Lett. {\bf 75}, 4512 (1995); {\bf 78}, 1972 (1997).
\bibitem{kosh}A.\ E.\ Koshelev, Phys.\ Rev.\ Lett. {\bf 77}, 3901 (1996).
\bibitem{kbmPRL98}A.\ E.\ Koshelev, L.\ N.\ Bulaevskii, and M.\ P.\ Maley, 
Phys. Rev. Lett. {\bf 81}, 9 (1998).
\bibitem{blk}L.\ N.\  Bulaevskii, M.\ Ledvij, and V.\ G.\ Kogan, Phys. Rev. B
{\bf 46}, 366, 11807 (1992).
\bibitem{Cubitt} S.\ L.\ Lee {\it et al.}, Phys.\ Rev.\ Lett.\ {\bf 71}, 3862 (1993); 
R.\ Cubitt {\it et al.}, Nature (London) {\bf 365}, 407 (1993).
\bibitem{sh}T. Shibauchi {\it et al.}, 
Phys.\ Rev.\ Lett.\ {\bf 83}, 1010 (1999) .
\bibitem{shib}T.\ Tamegai {\it et al.},
Advances in Superconductivity X, Proceedings of the 
10th International Symposium on Superconductivity (ISS '97), 
October 27-30, 1997, Gifu, Springer-Verlag Tokyo 1998. 
\bibitem{mats}Y.\ Matsuda, unpublished.
\bibitem{BlatterPRB96}  G.\ Blatter {\it et al.}, Phys.\ Rev.\ B {\bf 54}, 
72 (1996).
\bibitem{bdmb}  L.\ N.\ Bulaevskii {\it et al.},
Phys.  Rev.  B {\bf 54}, 7521 (1996).  
\bibitem{phasefluct} Near $T_{c}$ temperature dependence of $\omega
_{0}$ is strongly renormalized by phase fluctuations and thermally
induced vortex-antivortex pairs.  
\bibitem{kkPRB93} A.\ E.\ Koshelev and P.\ H.\ Kes, Phys.\ Rev.\ B
{\bf 48} 6539 (1993).
\bibitem{bm}L.\ N.\ Bulaevskii and M.\ P.\ Maley, SPIE Proceedings Series, v. 2157, 
16 (1994).
\bibitem{shm}G.\ S.\ Mkrtchyan and V.\ V.\ Shmidt, Zh.\ Eksp.\ Teor.\ Fiz.\ {\bf 61}, 
367 (1971) [Sov.\ Phys.\ JETP {\bf 34}, 195 (1972)].
\bibitem{yurgens}  A.\ Yurgens {\it et al.}, 
Phys.\ Rev.\ B {\bf 60} 12480 (1999).
\end{references}
\end{document}